\documentclass[10pt]{iopart}

\expandafter\let\csname equation*\endcsname\relax
\expandafter\let\csname endequation*\endcsname\relax

\usepackage{amsfonts,amsmath,amssymb}
\usepackage{hhline}
\usepackage{graphicx,psfrag,xcolor}
\usepackage{hyperref}
\usepackage{bm}
\definecolor{linkcolor}{HTML}{0176ba}
\definecolor{urlcolor}{HTML}{0176ba} 
\definecolor{citecolor}{HTML}{900020}
\hypersetup{pdfstartview=FitH,  linkcolor=linkcolor,urlcolor=urlcolor, citecolor=citecolor, colorlinks=true}

\begin{document}
\title[Multipole Born series approach to light scattering by Mie-resonant nanoparticle structures]{Multipole Born series approach to light scattering by Mie-resonant nanoparticle structures}

\author{Nikita A. Ustimenko$^1$, Danil F. Kornovan$^1$, Kseniia V. Baryshnikova$^1$, Andrey B. Evlyukhin$^{2,1}$, and Mihail I. Petrov$^1$}

\address{$^1$ School of Physics and Engineering, ITMO University, St. Petersburg 197101,  Russia}
\address{$^2$ Institute of  Quantum Optics, Leibniz Universitat Hannover, Hannover 30167, Germany}
\ead{nikita.ustimenko@metalab.ifmo.ru}

\begin{abstract}
Exciting optical effects such as polarization control, imaging, and holography were demonstrated at the nanoscale using the complex and irregular structures of nanoparticles with the multipole Mie-resonances in the optical range. The optical response of such particles can be simulated either by full wave numerical simulations or by the widely used analytical coupled multipole method (CMM), however, an analytical solution in the framework of CMM can be obtained only in a limited number of cases. In this paper, a modification of the CMM in the framework of the Born series and its applicability for simulation of light scattering by finite nanosphere structures, maintaining both dipole and quadrupole resonances, are investigated. The Born approximation simplifies an analytical consideration of various systems and helps shed light on physical processes ongoing in that systems. Using Mie theory and Green's functions approach, we analytically formulate the rigorous coupled dipole-quadrupole equations and their solution in the different-order Born approximations. We analyze in detail the resonant scattering by dielectric nanosphere structures such as dimer and ring to obtain the convergence conditions of the Born series and investigate how the physical characteristics such as absorption in particles, type of multipole resonance, and geometry of ensemble influence the convergence of Born series and its accuracy. 
\end{abstract}
\vspace{2pc}
\maketitle
\ioptwocol

\section{Introduction}

The rapidly developing all-dielectric nanophotonics~\cite{Kruk2017,Koshelev2020,babicheva2021multipole} brings new functionalities to nanoscale devices and systems for nonlinear generation~\cite{Wang2018,zograf2020highharmonic,shi2020progressive, Saerens2020, Frizyuk2021}, polarization control~\cite{Arbabi2015,Kruk2016}, sensing~\cite{Yesilkoy2019,Tseng2020}, lasing~\cite{Ha2018,Murai2020}, and imaging~\cite{Chen2020flat}. On this way, achieving high efficiency of future devices requires well-designed nanostructures with optimized parameters, which is often connected to resource-intensive simulations, especially when it comes to irregular structures such as holographic metasurfaces~\cite{Wang2016,Huang2018} and metalenses~\cite{Lin2014,Chen2019,Li2021}. Moreover, the extensively developing machine learning algorithms for inverse design of nanophotonic structures rely upon simulations of large parameter sets for proper training of neural networks~\cite{Yao2019,So2020}. Thus, fast and effective methods of the optical properties modeling are constantly required. In many cases high accuracy of numerical calculations is excessive, and approximate algorithms may be suitable~\cite{Yao2019}. In this paper, we focus on the Born series approach to modeling of optical response of ensembles of resonant subwavelength scatterers.          

\begin{figure}
    \centering
    \includegraphics[scale=0.9]{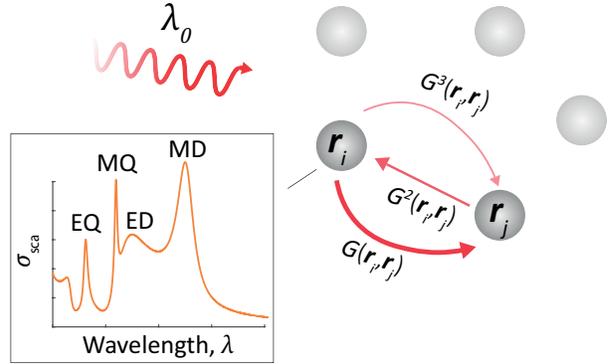}
    \caption{Light scattering by a group of nanoparticles at the conditions of their individual multipole resonances ($\lambda_0$ is the wavelength of an incident wave). If electromagnetic coupling between particles is quite weak the interaction in the system can be approximated by successive rescattering of secondary waves (multipole Born series approach). Each such rescattering between multipoles is described by the power of Green's function $G(\bm{r}_i, \bm{r}_j)$. Inset shows the scattering cross section $\sigma_{\rm sca}$ of an individual nanoparticle supporting magnetic dipole (MD), electric dipole (ED), electric quadrupole (EQ), and magnetic quadrupole (MQ) resonances. }
    \label{fig:born_expansion_scheme}
\end{figure}

The Born series formalism is a method for simulating wave propagation in many-body scattering problems. In nanophotonics, one can analyze the optical response of an ensemble of many nanoparticles in a perturbative manner (see Fig. \ref{fig:born_expansion_scheme})~\cite{Labani1990, Keller1993, fan2006application, singham1988light, Bereza2017, Babicheva2017, Ustimenko2021}.
This method is based on constructing a convergent Born series and replacing it with a finite sum that successively approximates the interaction between particles where the accuracy depends on the number of terms included in the sum (i.e., on the Born approximation order). Born approximations of different orders have been used to simulate tip-substrate interaction~\cite{Labani1990, Keller1993}, calculate polarizability of a non-spherical particle~\cite{fan2006application, singham1988light}, model the antireflective properties of nanoparticle coatings~\cite{Babicheva2017} and optimize metalens design~\cite{Ustimenko2021}. The applicability of the Born series method, as well as its convergence, are determined by the strength of electromagnetic coupling in the system. To the best of our knowledge, there is no exhaustive physical analysis of Born series applicability to simulate Mie-resonant nanostructures~\cite{evlyukhin2012demonstration, Kuznetsov2016} with qualitative criteria of the series convergence. In this work, we try to give such analysis and criteria and compare the Born series approach to the coupled multipole method (CMM)~\cite{babicheva2021multipole, Purcell1973, Bohren1983, Evlyukhin2010, Babicheva2019}.

The paper is organized as follows: in Section~\ref{sec:II}, we overview a general scheme of Born series with respect to multipole scattering model; in Section~\ref{sec:dimer}, we in details investigate the Born series convergence for a nanosphere dimer supporting Mie resonances and derive simple yet helpful  analytical results; Section~\ref{sec:ring} discusses the extension of the consideration to the case of particle ring. Finally, in Section~\ref{sec:timing}, we study the computational efficiency of Born approximation of various orders and compare their productivity with the CMM approach. 

\section{Multipole Born series method}
\label{sec:II}
At the center of our consideration is the problem of electromagnetic plane wave $\bm{E}_0 e^{\mathrm{i}\bm{k}\bm{r} - \mathrm{i}\omega t}$ scattering by finite-sized systems of $N$ Mie-resonant spherical nanoparticles with the dielectric permittivity $\varepsilon(\omega)=\varepsilon^{\prime}(\omega)+\mathrm{i}\varepsilon^{\prime\prime}(\omega)$ embedded in a vacuum as shown in Figure \ref{fig:born_expansion_scheme}. In the optical range, nanoparticles with a diameter of several hundred nanometers, made of a high-index dielectric or semiconductor, maintain dipole and quadrupole resonances of both electric and magnetic types~\cite{evlyukhin2012demonstration,Kuznetsov2016,Kuznetsov2012} (see inset in Fig. \ref{fig:born_expansion_scheme}). The optical response of such nanoparticle ensembles can be described by taking into account only the contribution of dipole and quadrupole modes while the contribution of higher-order multipoles is insignificant. The coupled multipole model (CMM)~\cite{Purcell1973, Babicheva2019} is used to describe electromagnetic interaction between multipoles generated in nanoparticles. 
In this model, for nanoparticle with number $j$ and position $\bm{r}_j$, the vector of electric dipole (ED) moment $\bm{p}^{j}$, the vector of magnetic dipole (MD) moment $\bm{m}^{j}$, the tensor of electric quadrupole (EQ) moment $\hat{Q}^{j}$, and the tensor of magnetic quadrupole (MQ) moment $\hat{M}^{j}$ are determined by the local electric $\bm{E}_{\mathrm{loc}}$ or magnetic $\bm{H}_{\mathrm{loc}}$ fields: 
\begin{gather}
\label{multipoles_defs}
\begin{gathered}
\bm{p}^{j} = \alpha_{p} \bm{E}_{\mathrm{loc}}\left(\bm{r}_j\right), \\
\bm{m}^{j} = \alpha_{m} \bm{H}_{\mathrm{loc}}\left(\bm{r}_j\right), \\
\hat{Q}^{j} = \frac{\alpha_{Q}}{2}\left[\bm{\nabla}_j\bm{E}_{\mathrm{loc}}\left(\bm{r}_j\right) + \bm{E}_{\mathrm{loc}}\left(\bm{r}_j\right) \bm{\nabla}_j\right], \\
\hat{M}^{j} = \frac{\alpha_{M}}{2}\left[\bm{\nabla}_j\bm{H}_{\mathrm{loc}}\left(\bm{r}_j\right) + \bm{H}_{\mathrm{loc}}\left(\bm{r}_j\right) \bm{\nabla}_j\right],
\end{gathered}
\end{gather}
where index $j = 1...N$, $\bm{\nabla}_j$ is the nabla operator with respect to $\bm{r}_j$; $\alpha_{p}$, $\alpha_{m}$, $\alpha_{Q}$, and $\alpha_{M}$ are the ED, MD, EQ, and MQ polarizabilities of a dielectric sphere, respectively [see Eqs. (\ref{polarizabilities}) in~\ref{sec:polarizabilities}]. 
A tensor $(\bm{\nabla} \bm{F} +  \bm{F}\bm{\nabla})$ is defined as follows:
\begin{gather*}
 (\bm{\nabla} \bm{F} +  \bm{F}\bm{\nabla})_{\beta \gamma} = \frac{\partial F_{\beta}}{\partial\gamma} + \frac{\partial F_{\gamma}}{\partial \beta},
\end{gather*}
where $\bm{F}$ is the vector of electric or magnetic field, indices $\beta = x,y,z$ and $\gamma = x,y,z$.

The local field acting on the $j$-th nanoparticle is composed of the external field and fields of multipoles generated in all other nanoparticles except the multipoles $j$-th nanoparticle:
\numparts
\begin{eqnarray}
\bm{E}_{\mathrm{loc}}\left(\bm{r}_j\right)&=& \bm{E}_{\mathrm{inc}}\left(\bm{r}_j\right) + {\bm{E}}_{p}'\left(\bm{r}_{j}\right)+{\bm{E}}_{m}'\left(\bm{r}_{j}\right)\nonumber\\
&&+ {\bm{E}}_{Q}'\left(\bm{r}_{j}\right)+{\bm{E}}_{M}'\left(\bm{r}_{j}\right), \label{el_field_loc}\\
\bm{H}_{\mathrm{loc}}\left(\bm{r}_j\right)& =& \bm{H}_{\mathrm{inc}}\left(\bm{r}_j\right) + {\bm{H}}_{p}'\left(\bm{r}_{j}\right)+{\bm{H}}_{m}'\left(\bm{r}_{j}\right) \nonumber\\
&&+ {\bm{H}}_{Q}'\left(\bm{r}_{j}\right)+{\bm{H}}_{M}'\left(\bm{r}_{j}\right),
\label{mag_field_loc}
\end{eqnarray}
\endnumparts
where ${\bm{E}}_{p}'\left(\bm{r}_{j}\right)$ is expressed through the all EDs generated in all nanoparticles except the $j$-th one (this is highlighted by $'$), and so on. The expressions for scattered electric $\bm{E}'$ and magnetic $\bm{H}'$ fields of the multipoles are provided in~\ref{sec:Multipole_fields} [see Eqs. (\ref{eq:multipole_fields})]. Thus, the multipole moments of a certain nanoparticle (\ref{multipoles_defs}) linearly depend on the multipole moments of other nanoparticles. Hence, to calculate the multipole moments of all nanoparticles arranged in the finite nanoparticle array, a system of linear equations should be solved:
\begin{gather}
\label{eq:cmm_system_matrix}
    \mathbf{Y} = \mathbf{Y}_0 + \hat{\mathbf{V}}\mathbf{Y}.
\end{gather}
Here $\mathbf{Y}$ is the supervector of the coupled multipole moments (\ref{multipoles_defs}), taking into account the interaction of particles. $\mathbf{Y}_0$ is the supervector of multipole moments excited only by the external wave, i.e., the local fields are replaced by the incident ones in (\ref{multipoles_defs}). The block matrix $\hat{\mathbf{V}}$ descibes the interaction between multipoles. The explicit forms of vectors $\mathbf{Y}$ and $\mathbf{Y}_0$, and matrix $\hat{\mathbf{V}}$ are added in~\ref{sec:system}.

Self-consisted CMM solution of Eq. (\ref{eq:cmm_system_matrix}) can be written as follows:
\begin{gather}
\label{eq:cmm_solution_matrix}
\mathbf{Y} = (\hat{\mathbf{I}} - \hat{\mathbf{V}})^{-1} \mathbf{Y}_0,    
\end{gather}
where $\hat{\mathbf{I}}$ is the corresponding identity matrix. The Born series expansion of (\ref{eq:cmm_solution_matrix}) is a matrix $(\hat{\mathbf{I}} - \hat{\mathbf{V}})^{-1}$ expansion in terms of powers of the matrix $\hat{\mathbf{V}}$:
\begin{gather}
\label{eq:born_series_matrix}
\mathbf{Y} = \hat{\mathbf{I}}\mathbf{Y}_0 + \hat{\mathbf{V}}\mathbf{Y}_0 + \hat{\mathbf{V}}^2\mathbf{Y}_0 + \hat{\mathbf{V}}^3\mathbf{Y}_0 + \hdots .   
\end{gather}
Replacing the series (\ref{eq:born_series_matrix}) by a finite sum, we can obtain the solution of (\ref{eq:cmm_system_matrix}) in the Born approximation. In the zero-order Born approximation, the interaction between multipoles is neglected:
\begin{gather}
\mathbf{Y} = \mathbf{Y}_0.   
\end{gather}
The $m$-th order Born approximation is expressed through the ($m-1$)-th order Born approximation:
\begin{gather}
\mathbf{Y}_m = \mathbf{Y}_0 + \hat{\mathbf{V}}\mathbf{Y}_{m-1}.  
\end{gather}

The main criterion of applicability of the Born series is its  convergence. The necessary convergence condition of the series (\ref{eq:born_series_matrix}) is $\mathrm{det}(\hat{\mathbf{I}} - \hat{\mathbf{V}}) \neq 0$; otherwise, a system should be far from the condition of configurational resonance~\cite{Keller1993}. In this case, a strong electromagnetic coupling between multipoles cannot be approximated by the Born approximation of any order. The sufficient condition is $\|\hat{\mathbf{V}}\| < 1$. Generally, the series (\ref{eq:born_series_matrix}) converges only when all eigenvalues of matrix $\hat{\mathbf{V}}$ are inscribed in a unit circle on a complex plane. This condition is mathematically strict but cannot be applied immediately to answer the question about the Born series convergence. From a physical point of view, the Born series diverges when a strong electromagnetic interaction between the nanoparticles appears. Obviously, at the Mie resonance inherent for each particle in the ensemble electromagnetic coupling is enhanced and a question of the Born series becomes very important.

\section{Nanosphere dimer}
\label{sec:dimer}
\begin{figure}
    \centering
    \includegraphics[scale=0.65]{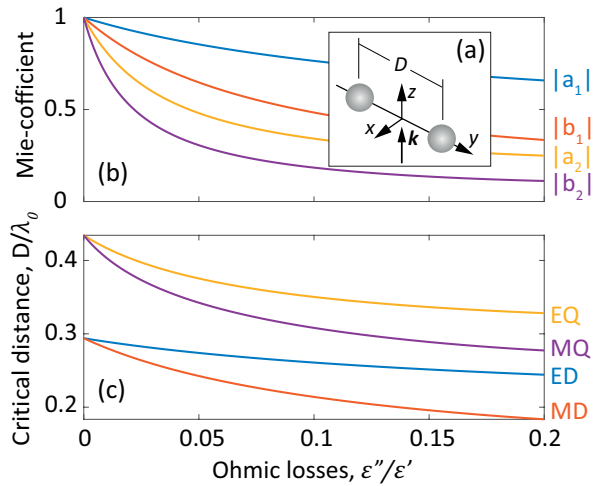}
    \caption{(a) Schematic of identical nanosphere dimer embedded in a vacuum with the marked wavevector of the external incident wave. (b) The resonant absolute value of Mie-coefficients as function of particle material $\varepsilon^{\prime\prime}$ (where $\varepsilon^{\prime} = 12.5$, and diameter of nanoparticle is 200 nm). (c) Critical center-center distance between specific multipoles in the dimer as function of particles $\varepsilon^{\prime\prime}$ for the considered isolated multipole resonances. The vector of electric field in external wave $\bm{E}_0 \parallel y$ for ED and EQ resonances, and $\bm{E}_0 \parallel x$ for MD and MQ resonances.
    }
    \label{fig:abs_Mie_Dcritical_vs_im_eps}
\end{figure}

We start our analysis from a dimer of identical nanospheres shown in Fig. \ref{fig:abs_Mie_Dcritical_vs_im_eps}(a). On the one hand, dimer structures  can be utilized for various nanophotonic purposes such as local magnetic field enhancement \cite{Bakker2015}, nonlinear generation enhancement \cite{Saerens2020, Frizyuk2021, Renaut2019}, controlling the directionality of scattering and others, and, on the other hand, offer  a simple analytical solution in the framework of the CMM.

Close to a particular multipolar resonance, one can consider the dominant contribution of the isolated (specific) resonant multipole. In this case, for the ED resonance, Eq. (\ref{eq:cmm_system_matrix}) can be written as follows:
\begin{gather}
\label{eq:cdm_system_dimer}
\begin{array}{lr}
  \bm{p}^1 = \alpha_p \bm{E}_0 +  \alpha_p k^2\varepsilon_0^{-1}\hat{G}^{pp}_{12} \bm{p}^{2}, \\
  \bm{p}^2 = \alpha_p\bm{E}_0 +  \alpha_p k^2\varepsilon_0^{-1}\hat{G}^{pp}_{21} \bm{p}^{1},
\end{array}
\end{gather}
where $\bm{p}^1$ and $\bm{p}^2$ are the dipole moments of the first and second nanoparticle, respectively, $k = 2\pi/\lambda$ is the free-space wavenumber of external plane wave, $\hat{G}^{pp}_{12} \equiv \hat{G}^{pp}(\bm{r}_1,\bm{r}_2)$ is the dyadic dipole Green's function having a symmetry property $\hat{G}^{pp}_{12} = \hat{G}^{pp}_{21}$ [see Eq. (\ref{dipole_green_formula}) in~\ref{section:green}]. The nanoparticles are separated from each other by the distance $D$ along the $y$-axis as shown in Fig. \ref{fig:abs_Mie_Dcritical_vs_im_eps}(a). 

Due to the structure symmetry, both dipole moments have only one non-zero component along the incident field:
\begin{gather}
\label{eq:dimer_cdm_solution}
    p^j_{\beta} = \frac{\alpha_p E_0}{1 - \alpha_p k^2 \varepsilon_0^{-1} G^{pp}_{12, \beta\beta}}, \quad j=1,2.
\end{gather}
where $\hat{G}^{pp}_{12, \beta\beta} = G^{pp}_{\beta\beta}(\bm{r}_1, \bm{r}_2)$ is the $\beta\beta$-element of Green's function $\hat{G}^{pp}_{12}$: $\beta = x$ for the transverse-polarized dipole moments ($\bm{E}_0 \parallel x$), and $\beta = y$ for the longitudinal polarization ($\bm{E}_0 \parallel y$). 


\begin{table}
(a) Convergence criteria\\
\begin{tabular}{|p{0.6cm}|p{3.2cm}|p{3.2cm}|  }
 \hline
 R/P& $\bm{E}_0 \parallel y$ & $\bm{E}_0 \parallel x$\\ 
 \hline
ED &  $6\pi |a_1|| k^{-1} G^{pp}_{12, yy}| < 1$ & $6\pi |a_1|| k^{-1} G^{pp}_{12, xx}| < 1$ \\
 \hline
 MD & $6\pi |b_1|| k^{-1} G^{pp}_{12, xx}| < 1$ & $6\pi |b_1|| k^{-1} G^{pp}_{12, yy}| < 1$ \\
 \hline
EQ & $60\pi |a_2| |k^{-3} B^{QQ}_{12, yz}| < 1$ & $60\pi |a_2| |k^{-3} B^{QQ}_{12, xz}| < 1$ \\
 \hline
MQ & $60\pi |b_2| |k^{-3} B^{QQ}_{12, xz}| < 1$ & $60\pi |b_2| |k^{-3} B^{QQ}_{12, yz}| < 1$ \\
 \hline
\end{tabular}
\\
(b) Critical distances ($\varepsilon^{\prime \prime}$ = 0)\\
\begin{tabular}{|p{0.6cm}|p{3.2cm}|p{3.2cm}|  }
 \hline
R/P& $\bm{E}_0 \parallel y$ & $\bm{E}_0 \parallel x$\\
 \hline
ED & $D > 0.29\lambda_{\mathrm{ED}}$ & $D > 0.21\lambda_{\mathrm{ED}}$ \\
 \hline
 MD & $D > 0.21\lambda_{\mathrm{MD}}$ & $D > 0.29\lambda_{\mathrm{MD}}$ \\
 \hline
EQ & $D > 0.44\lambda_{\mathrm{EQ}}$ & $D > 0.33\lambda_{\mathrm{EQ}}$ \\
 \hline
MQ & $D > 0.33\lambda_{\mathrm{MQ}}$ & $D > 0.44\lambda_{\mathrm{MQ}}$ \\
 \hline
\end{tabular}
\caption{Conditions and parameters of Born series convergence for the dimer of ED, MD, EQ, and MQ scatterers in vicinity of isolated resonance (R) for both linear polarizations (P) of incident wave. (a) Summary of convergence criteria. (b) Solutions of inequalities in (a) for Mie resonances in non-absorptive particles ($\varepsilon''=0$). Here $\lambda_{\rm ED}$, $\lambda_{\rm MD}$, $\lambda_{\rm EQ}$, $\lambda_{\rm MQ}$ are the wavelengths of ED, MD, EQ, MQ resonances, respectively; $a_n$ and $b_n$ are the Mie-coefficients; $G^{pp}_{12, xx}$ and $G^{pp}_{12, xx}$ are the elements of dipole Green's function $\hat{G}^{pp}(\bm{r}_1, \bm{r}_2)$ [see Eq. (\ref{dipole_green_formula}) in~\ref{section:green}]; expressions for quantities $B^{QQ}_{12, xz}$ and $B^{QQ}_{12, yz}$ are written in~\ref{sec:solutions} [see Eq. (\ref{eq:definition_BQQ})].}
\label{convergence_table}
\end{table}

The dipole moments calculated by Eq. (\ref{eq:dimer_cdm_solution}) take into account the electromagnetic coupling between two dipoles rigorously. We can also approximate the coupling by expanding the moments (\ref{eq:dimer_cdm_solution}) into the Born series. For this aim, we expand the denominator in (\ref{eq:dimer_cdm_solution}) into geometric series:
\begin{gather}
\label{eq:dimer_born_series}
p^j_{\beta} = \alpha_p E_0\sum\limits_{s=0}^{\infty}\left(\alpha_p k^2 \varepsilon_0^{-1} G^{pp}_{12, \beta\beta}\right)^s, \quad j=1,2.
\end{gather}
The solution of Eqs. (\ref{eq:cdm_system_dimer}) in the Born approximation of $m$-th order is given by transition from an infinite series to a finite sum in Eq. (\ref{eq:dimer_born_series}), i.e., $\sum\limits_{s=0}^{\infty} \rightarrow \sum\limits_{s=0}^{m}$.
The convergence criteria of geometric series such as Eq. (\ref{eq:dimer_born_series}) is well-known~\cite{Sveshnokov1978}:
\begin{gather}
\label{eq:convergence_condition}
    |\alpha_p \cdot k^2 \varepsilon_0^{-1} G^{pp}_{12,\beta\beta}| < 1.
\end{gather}
The ED polarizability of a homogeneous nanosphere can be expressed through the Mie-coefficient $a_1(nkR)$ \cite{Bohren1983, Evlyukhin2010}, depending on the particle refractive index $n$ and radius $R$, and incident field wavelength,
then the convergence criterion for the \textit{isolated} ED resonance can be formulated as follows:
\begin{gather}
\label{eq:convergence_condition_a1}
6\pi \cdot |a_1| \cdot |k^{-1} G^{pp}_{12, \beta\beta}| < 1.
\end{gather}
For other \textit{isolated} multipole resonances including MD, EQ, and MQ resonances, the convergence criteria are written in Table~\ref{convergence_table}(a). The MD, EQ, and MQ polarizabilities of spherical particle in the framework of Mie-theory are written in~\ref{sec:polarizabilities} [see Eqs. (\ref{polarizabilities})] while the solutions of Eq.~(\ref{eq:cmm_system_matrix}) for the dimers of such multipole scatterers are provided in~\ref{sec:solutions}.

In the case of non-absorbing particles ($\varepsilon'' = 0$), the Mie-resonance condition for electric or magnetic modes provides~\cite{Hulst1981} $a_n = 1$ or $b_n=1$ correspondingly, which immediately provides us with the at-resonance convergence criteria of the multipole Born series. Solving the  inequality (\ref{eq:convergence_condition_a1}), we can find the \textit{critical} distances which, for the longitudinal and transverse ED modes, are $D^{\rm (L)} = 0.29\lambda_0$ and $D^{\rm (T)} = 0.21\lambda_0$, respectively. If the distance between the centers of
particles is larger than the critical one, the Born series converges, otherwise, it diverges. The critical distances for other multipole resonances are specified in Table~\ref{convergence_table}(b). 

Once the non-zero Ohmic losses are present  ($\varepsilon'' > 0$), the Mie-coefficients at the resonances become less than unity $|a_n|,\ |b_n|<1$ that decreases the critical distances. Fig.~ \ref{fig:abs_Mie_Dcritical_vs_im_eps}(b) shows the at-resonance absolute values of the four Mie-coefficients for varying $\varepsilon''$ of nanoparticle material.  In the simulation, $\varepsilon'$ of the particle remained constant while the particle diameter $d$ was changed in order to keep the resonant wavelength constant for different $\varepsilon''$. Since the convergence criteria depend on the absolute values of Mie-coefficients [see Table~\ref{convergence_table}(a)], with increasing of materials losses the critical distances also decrease for all multipole resonances [see Fig.~\ref{fig:abs_Mie_Dcritical_vs_im_eps}(c)]. This behavior has a clear physical explanation as additional losses suppress nanoparticle scattering making electromagnetic coupling between two particles weaker, therefore, improving the convergence of the Born series.


\begin{figure}
    \centering
    \includegraphics[scale=0.5]{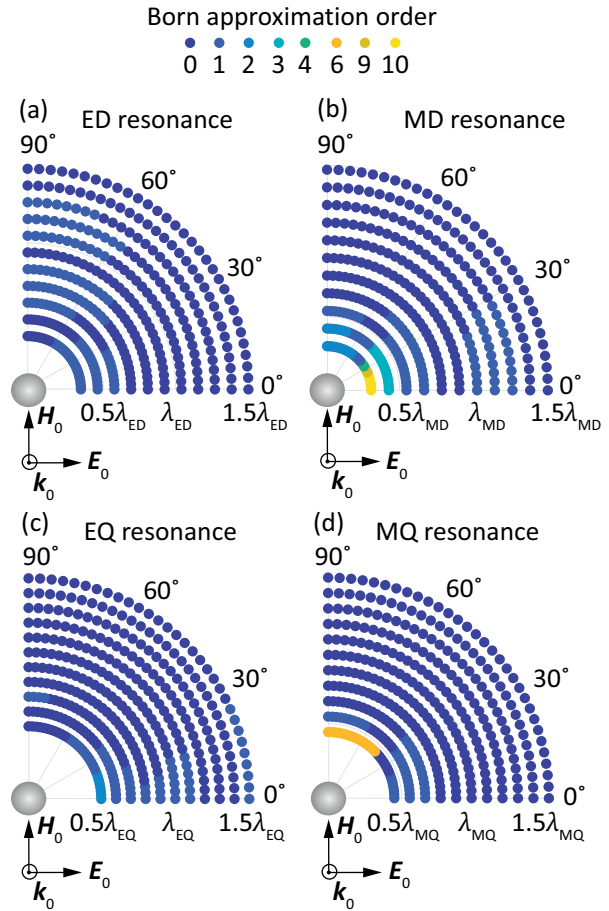}
    \caption{The Born approximation order as a function of the second particle position while the first particle is placed at the coordinate system origin for the (a) ED resonance ($\lambda_{\rm ED} = 555$ nm), (b) MD resonance ($\lambda_{\rm MD} = 731$ nm), (c) EQ resonance ($\lambda_{\rm EQ} = 407$ nm), and (d) MQ resonance ($\lambda_{\rm MQ} = 505$ nm). The position of the circle point indicates the second particle position, and its color corresponds to the order of Born approximation that allows calculating the extinction cross section with an error (\ref{eq:ECS_error}) of less than 10\%. The distance between the particles is measured in resonant wavelength $\lambda$. The particle diameter $d = 200$ nm, permittivity of the particle material $\varepsilon = 12.5$. The normally incident plane wave is horizontally polarized.}
    \label{fig:born_order_2d}
\end{figure}

Above we considered only the isolated resonances and took into account only the contributions of corresponding resonant multipoles. In the general case, the higher order multipoles can be excited simultaneously by a linearly polarized wave and their contribution requires larger distances between particles in order to obtain convergence: $D > 0.29\lambda_0$ for the dipole resonance, and $D > 0.44\lambda_0$ for the quadrupole resonance. For absorptive particles, the critical distances can be calculated using the inequalities from Table~\ref{convergence_table}(a).

The critical distance values provide us with the applicability conditions of the Born series approximation, however, they do not provide any information about its accuracy. For this reason, we analyzed the scattering cross section and compared it with the results obtained from CMM introducing the accuracy parameter $\Delta$: 
\begin{gather}
\label{eq:ECS_error}
   \Delta = \frac{|\sigma^{(m)} - \sigma^{(\mathrm{CMM})}|}{\sigma^{(\mathrm{CMM})}},
\end{gather}
where $\sigma^{(\mathrm{CMM})}$ and $\sigma^{(m)}$ are the scattering cross sections of the dimer calculated using the CMM multipole moments Eq.~(\ref{eq:cmm_solution_matrix}) and multipole moments in the $m$-th Born approximation, respectively. In the dipole-quadrupole approximation, the scattering cross section of non-absorptive particle structure in a vacuum can be calculated by the following formula~\cite{Babicheva2019}:
\begin{eqnarray}
\sigma =&&  \dfrac{k}{\varepsilon_0 |E_0|^2} \operatorname{Im}\sum\limits_{j=1}^{N} \left[ \bm{E}_0^{\ast}(\bm{r}_j) \cdot \bm{p}^j  +\mu_0 \tfrac{\left[\bm{\nabla}\bm{H}_0^{\ast}(\bm{r}_j)\right]^T}{2} : \hat{M}^j \right.\nonumber\\ &&\left. + \mu_0 \bm{H}_0^{\ast}(\bm{r}_j) \cdot \bm{m}^j+ \tfrac{\bm{\nabla}\bm{E}_0^{\ast}(\bm{r}_j) + \bm{E}_0^{\ast}(\bm{r}_j)\bm{\nabla}}{12} :  \hat{Q}^j \right],
\end{eqnarray}
the asterisk $^{\ast}$ denotes complex conjugation, $^T$ denotes the transpose operation, and the signs $\cdot$ and : denote the scalar products between vectors and dyads (tensors), respectively.

We placed one particle at the origin of the coordinate system, and varied the position of the second particle, while the distance between the particles was simultaneously greater than the critical parameter ($0.29\lambda_0$ or $0.44\lambda_0$) and their diameter (200 nm). Figure~\ref{fig:born_order_2d} shows the minimal number among the Born approximation orders $m$ one should take in order to provide the accuracy parameter error $\Delta\leq 0.1$ for considered resonances. In Fig.~\ref{fig:born_order_2d}, we already take into account  four multipoles (ED, MD, EQ, and MQ). One can see that the  convergence of the series at MD resonance is much weaker than for ED as the resonance is much more pronounced and the scattered partial fields are stronger. At the same time, the series convergence is slower along the E-field when the distance between the nanoparticles becomes small enough owing to stronger near-field interaction between particles. Very similar behavior is observed for quadrupole resonances in Figures~\ref{fig:born_order_2d}(c) and \ref{fig:born_order_2d}(d).   

\section{Nanosphere ring}
\label{sec:ring}
\begin{figure}
    \centering
    \includegraphics[scale=0.52]{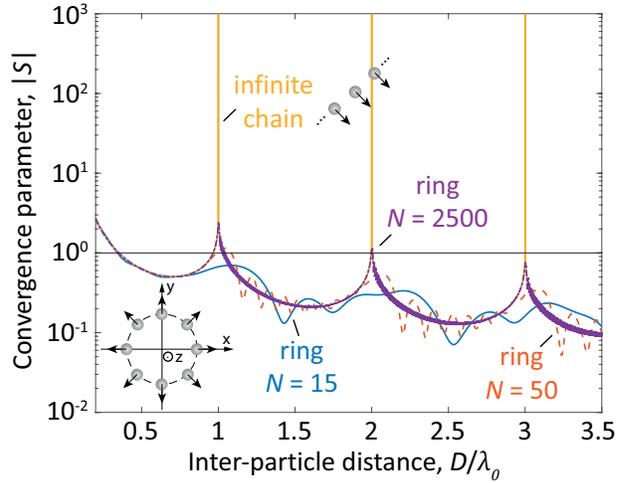}
    \caption{The convergence parameter $|S|$ for the radial mode of a ring as function of the inter-particle distance for the number of particle $N = 15$ (blue solid line), $N = 50$ (red dashed line), and $N = 2500$ (violet dashed-dotted line). Solid yellow line corresponds to the convergence parameter for the transversally polarized dipole mode of infinite chain. In lower inset: schematic of nanoparticle ring, and the radial ring dipole mode. In upper inset: schematic of infinite particle chain, and its transverse dipole mode.}
    \label{fig:ring structure}
\end{figure}

Another very illustrative ensemble geometry yet allowing rigorous analysis is a regular ring of $N$ nanoparticles shown in Figure~\ref{fig:ring structure}. 
The ring structures often become a building block of light focusing metasurface lenses ~\cite{Chen2020flat}, thus understanding their optical response can be critical for designing metalens structures. The interactions between the nanoparticles influence the convergence of multipole Born series close to the collective resonances. We consider the system (\ref{eq:cmm_system_matrix}) in purely ED approximation solve assuming and focusing on the mode with radial polarization of dipole moments (see Figure~\ref{fig:ring structure}), which  can be excited, for instance, by vectorial optical beams such as radially polarized Bessel beam \cite{Zhan2012}. The Eq. (\ref{eq:cmm_system_matrix}) in this case:
\begin{gather}
\label{eq:cmm_system_ring}
 \bm{p}^j_{\rho} = \alpha_p \bm{E}_0(\bm{r}_j) + \alpha_p \frac{k^2}{\varepsilon_0}\sum\limits_{l=1, l\neq j}^{N}G^{pp}_{\rho\rho}(\bm{r}_j,\bm{r}_l) \bm{p}^l_{\rho},
\end{gather}
where $\bm{r}_j = R[\cos{(\varphi_j)}, \sin{(\varphi_j)}, 0]^T$ is the position of the $j$-th nanoparticle in ring of radius $R$, $\varphi_j = 2\pi (j-1)/N$ is the angular coordinate of the $j$-th nanoparticle, $G^{pp}_{\rho\rho}$ is the component of the vacuum dipole Green’s tensor in cylindrical coordinates [see Eq.~(\ref{GFCYL2}) in~\ref{appendix:ring_dipole_sum}], and $\bm{p}^j_{\rho} = [p^j_x, p^j_y]^T$. Note that $p^j_z = 0$ for all nanoparticles in the structure.

Since the external field $\bm{E}_0(\bm{r}_j) = E_0[\cos{(\varphi_j)},$ $\sin{(\varphi_j)}, 0]^T$, the solution of (\ref{eq:cmm_system_ring}) can be found using the ansatz $\bm{p}^j_{\rho} = \tilde{p}[\cos{(\varphi_j)}, \sin{(\varphi_j)}]^T$. Inserting this anzats in Eq. (\ref{eq:cmm_system_ring}), we obtain the expression for $\tilde{p}$:
\begin{gather}
\label{eq:ring_cmm_solution}
   \tilde{p} = \underbrace{\frac{\alpha_p}{1 - \alpha_p k^2 \varepsilon_0^{-1} \tilde{G}^{pp}_{\rho\rho}}}_{\alpha_{\mathrm{eff}}}E_{0},
\end{gather}
where the effective polarizability of the ring $\alpha_{\mathrm{eff}}$ is introduced, $\tilde{G}^{pp}_{\rho\rho}$ is the dipole (lattice) sum of the radial ring mode [see Eq.~(\ref{eq:G0}) in~\ref{appendix:ring_dipole_sum}].

The Born series for the ring dipole moments is similar to the one for dimer (\ref{eq:dimer_born_series}), hence, the convergence criterion for the ring is following:
\begin{gather}
\label{eq:ring_convergence_parameter}
|S|= |\alpha_p k^2 \varepsilon_0^{-1} \tilde{G}^{pp}_{\rho\rho}| < 1,    
\end{gather}
while at the configurational resonance, when $|S| = 1$, the Born series always diverges. Figure~\ref{fig:ring structure} shows the convergence parameter $|S|$ [defined in (\ref{eq:ring_convergence_parameter})] for the radial mode of rings with different inter-particle distances and radii. It can be seen that the Born series may diverge  when the distance between particles is very small or when the distance between the dipoles is equal to an integer number of resonant wavelengths. The latter case corresponds to the geometric (diffraction) resonances indicating a strong electromagnetic interaction between nanoparticles. Also, increasing the number of radially oriented dipoles in the ring, the convergence parameter for the considered mode of ring approaches the dependence for a transverse mode of a linear chain  (see Fig.~\ref{fig:ring structure}). Indeed, in the limit of an infinite ring $N,R \to \infty$ with fixed inter-particle distance, the dipole sum of a ring tends to the dipole sum of a chain known analytically (see~\ref{appendix:ring_dipole_sum}).


The convergence parameter $|S|$ of the mode can be tied with its dispersion and $Q$-factor. The polarizability of a single particle in the vicinity of isolated high-$Q$ resonance $\omega_0$ can be represented as:
\begin{gather}
\label{eq:alpha_single}
\alpha_p = -\alpha_0 \frac{\gamma_0/2}{\omega - \omega_0 + \mathrm{i}\gamma_0/2},   
\end{gather}
where $\alpha_0 = 6\pi\varepsilon_0 k_0^{-3}$ is the resonant value of dipole polarizability, $\gamma_0/2$ is the radiative losses rate of a single particle. Inserting (\ref{eq:alpha_single}) into (\ref{eq:ring_cmm_solution}) and using quasi-resonant approximation $\tilde{G}^{pp}_{\rho \rho}(\omega) \approx \tilde{G}^{pp}_{\rho \rho}(\omega_0)$, we obtain the following expression for the effective polarizability:
$\alpha_{\mathrm{eff}} \approx -\alpha_0 \frac{\gamma_0/2}{\omega - \omega_0 -  \Delta \omega_0 + \mathrm{i}\gamma/2}$, where $\Delta \omega_0$ is then detuning of resonance frequency from the resonant of an individual particle, $\gamma/2$ is the radiative losses rate of the ring mode. At the resonance ($\omega = \omega_0$) of the single particle, the convergence condition (\ref{eq:ring_convergence_parameter}) can be written as:
\begin{gather}
\label{eq:convergence_dispersion}
    |S| = \sqrt{\left(\frac{\Delta \omega_0}{\gamma_0/2}\right)^2 + \left(\frac{\gamma}{\gamma_0} - 1\right)^2} < 1.
\end{gather}
If $|\Delta \omega_0| \ll \gamma_0/2$ inequality (\ref{eq:convergence_dispersion} can be expressed through the ratio of $Q$-factors:
\begin{gather}
  |S| \approx \left|\frac{Q_0}{Q} - 1\right| < 1, 
\end{gather}
where $Q_0$ is the total $Q$-factor of a single-particle resonance, and $Q$ the is $Q$-factor of the ring mode, showing that, indeed, the stronger is the collective resonance the slower is the convergence of the Born series as for $Q\gg Q_0$ $|S|\sim 1$. 

\section{The performance of Born series method}
\label{sec:timing}
In the last part of our paper, we turn to analyze the performance efficiency of the Born approximation. We compare the computational time required for a rigorous solution of the system (\ref{eq:cmm_system_matrix}) with the solution of (\ref{eq:cmm_system_matrix}) in the Born approximations for both dipole and dipole-quadrupole models. We consider a ring of $N$ spherical nanoparticles (of diameter $d$ = 200 nm) with the inter-particle distance $D = 0.6 \lambda_{\rm ED}$ such that the Born series converges at the wavelength of electric dipole resonance $\lambda_{\rm ED} = 615$ nm. For the ring structure of $N$ nanoparticles, a dimension of the matrix of the system (\ref{eq:cmm_system_matrix}) is $6N \times 6N$ and $24N \times 24N$ for dipole and dipole-quadrupole models, respectively. We vary the particle number $N$ from 20 to  120 and estimate the time required to compute the solution with help of \textit{linsolve} function in Matlab for coupled dipoles (multipoles)  and the time required for Born series summation. 

\begin{figure}
    \centering
    \includegraphics[scale=0.5]{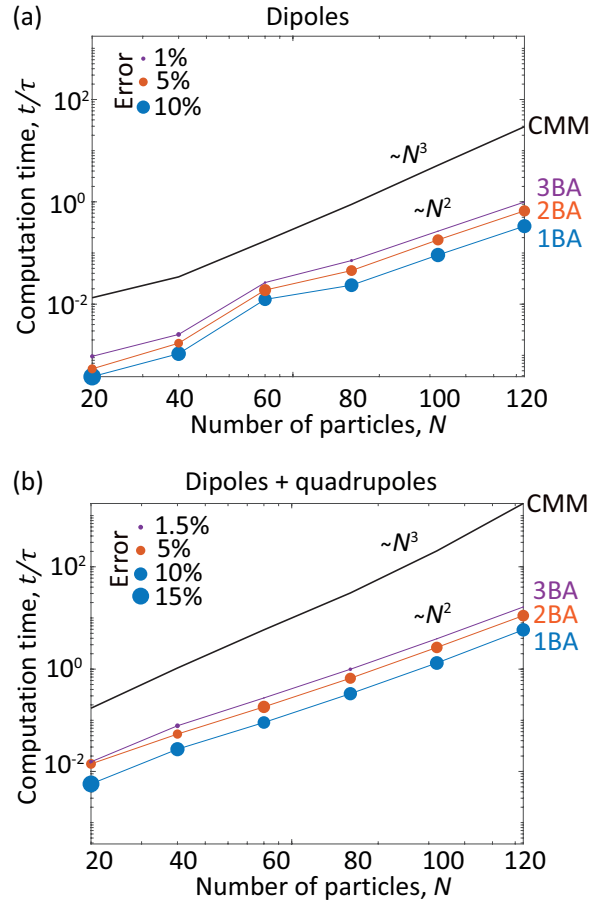}
    \caption{Normalized computation time required for solving the  system (\ref{eq:cmm_system_matrix}) in the coupled dipole (a) and coupled dipole-quadrupole (b) models and using the Born series approach (color circles and lines) as function of particles number. The size of circle is proportional to the magnitude of the relative error $\Delta$ of the scattering cross section calculated in the Born approximation and compared to CMM.} 
    \label{fig:calc_time_and_sigma_error_vs_N}
\end{figure}

The results of the comparison are presented in Figure~\ref{fig:calc_time_and_sigma_error_vs_N} in normalized units. In Fig.~\ref{fig:calc_time_and_sigma_error_vs_N}, $\tau$ is the average time of \textit{linsolve} work for the $1000\times1000$ matrix and $1000\times1$ right-hand side column of random complex numbers. All quantities in Fig.~\ref{fig:calc_time_and_sigma_error_vs_N} were averaged after ten iterations. The computation time of CMM behaves as $O(N^{3})$ while the Born series computation time behaves as $O(N^{2})$, which  provides almost one and two orders of magnitude decrease of computation time comparing to the coupled dipole and dipole-quadrupole solutions for $N=100$ particles. At the same time, the numerical error $\Delta$, provided by the Eq.~(\ref{eq:ECS_error}) and indicated by the circle diameter, stays well below 2\% for the third-order Born approximation. 

\section{Conclusion}\label{section:conclusion}  
In this paper, we analyzed the Born series approach for modeling the optical response of finite arrays of Mie-resonant nanoparticles. The Born series method allows us to approximately compute the interaction in particle systems that significantly reduce the total computation time and usage of computational resources comparing to the rigorous coupled multipole models. We investigate the accuracy of this method and, in particular, a convergence of the Born series in the vicinity of Mie-resonances. Under the dipole and quadrupole approximation, we analytically and numerically find the convergence conditions for the nanosphere dimer and nanosphere ring. We believe that the proposed simulation method and obtained results will be useful for the optimization and modeling of nanophotonic systems such as metasurfaces, metalenses, and nanoantennas.  

\ack{The work was supported by Russian Science Foundation (project 21-79-10190), and Deutsche Forschungsgemeinschaft (DFG, German Research Foundation) under Germany’s Excellence Strategy within the Cluster of Excellence PhoenixD (EXC 2122, Project ID 390833453).}

\appendix
\section{Mie-polarizabilities of spherical particle}\label{sec:polarizabilities}
For Mie scattering regime, the dipole and quadrupole polarizabilities of spherical nanoparticles in the field of external plane wave are expressed through the Mie-coefficients $a_n$ and $b_n$~\cite{Bohren1983,Babicheva2019}:
\begin{align}
\label{polarizabilities}
\begin{aligned}
    &\alpha_{p}=\mathrm{i} \frac{6 \pi \varepsilon_{0}\varepsilon_S}{k_S^{3}} a_{1}, \quad &
    &\alpha_{m}=\mathrm{i} \frac{6 \pi}{k_S^{3}} b_{1}, \\
    &\alpha_{Q}=\mathrm{i} \frac{120 \pi \varepsilon_{0}\varepsilon_S}{k_S^{5}} a_{2}, \quad &
    &\alpha_{M}=\mathrm{i} \frac{40 \pi}{k_S^{5}} b_{2},
\end{aligned}
\end{align}
where i is the imaginary unit, $\varepsilon_{0}$ is the dielectric permitivitty of vacuum, $\varepsilon_S$ is the dielectric permitivitty of particles surrounding medium, $k_S = k \sqrt{\varepsilon_S}$ is the wavenumber in medium with permitivitty $\varepsilon_S$. In the main text, $\varepsilon_S = 1$.

\section{Multipole fields}\label{sec:Multipole_fields}
Electric and magnetic fields of dipoles and qudrupoles at the nanoparticle point $\bm{r}_j$ generated by nanoparticles located at the points $\bm{r}_l$ ($l \neq j$)~\cite{Babicheva2019}:
\begin{align}
\label{eq:multipole_fields}
\begin{aligned}
{\bm{E}}_{p}'\left(\bm{r}_{j}\right) &= \frac{k^{2}}{\varepsilon_{0}} \sum\limits_{l = 1, l \neq j}^{N} \hat{G}_{j l}^{pp} \bm{p}^l,\\
{\bm{H}}_{p}'\left(\bm{r}_{j}\right) &= \frac{c k}{\mathrm{i}} \sum\limits_{l = 1, l \neq j}^{N} \hat{G}_{j l}^{pm} \bm{p}^{l}, \\
{\bm{E}}_{m}'\left(\bm{r}_{j}\right) &=
\frac{\mathrm{i} k}{c \varepsilon_{0}} \sum\limits_{l = 1, l \neq j}^{N} \hat{G}_{j l}^{pm} \bm{m}^{l}, \\
{\bm{H}}_{m}'\left(\bm{r}_{j}\right) &= k_S^{2} \sum\limits_{l = 1, l \neq j}^{N} \hat{G}_{j l}^{pp} \bm{m}^{l}, \\
\end{aligned}
\end{align}
\begin{align*}
\begin{aligned}
{\bm{E}}_{Q}'\left(\bm{r}_{j}\right) &= \frac{k^{2}}{\varepsilon_{0}} \sum\limits_{l = 1, l \neq j}^{N} \hat{G}_{j l}^{QQ}\left(\hat{Q}^{l} \bm{n}_{l j}\right),\\
{\bm{H}}_{Q}'\left(\bm{r}_{j}\right) &= \frac{c k}{\mathrm{i}} \sum\limits_{l = 1, l \neq j}^{N} \hat{G}_{j l}^{QM} \left(\hat{Q}^{l} \bm{n}_{l j}\right), \\
{\bm{E}}_{M}'\left(\bm{r}_{j}\right) &= 3 \frac{\mathrm{i} k}{c \varepsilon_{0}} \sum\limits_{l = 1, l \neq j}^{N} \hat{G}_{j l}^{QM} \left(\hat{M}^{l} \bm{n}_{l j}\right),\\ 
{\bm{H}}_{M}'\left(\bm{r}_{j}\right) &= 3 k_S^{2} \sum\limits_{l = 1, l \neq j}^{N} \hat{G}_{j l}^{QQ} \left(\hat{M}^{l} \bm{n}_{l j}\right), 
\end{aligned}
\end{align*}
where $c$ is the light speed in vacuum, $\bm{n}_{l j}$ is the unit vector from $j$-th nanoparticle to $l$-th nanoparticle: $\bm{n}_{l j} = (\bm{r}_{j} - \bm{r}_{l})/|\bm{r}_{j} - \bm{r}_{l}|$.  $\hat{G}_{j l}^{pp} = \hat{G}^{pp}(\bm{r}_j, \bm{r}_l)$, $\hat{G}_{j l}^{pm} = \hat{G}^{pm}(\bm{r}_j, \bm{r}_l)$, $\hat{G}_{j l}^{QQ} = \hat{G}^{QQ}(\bm{r}_j, \bm{r}_l)$, $\hat{G}_{j l}^{QM} = \hat{G}^{QM}(\bm{r}_j, \bm{r}_l)$ are the dyadic Green's functions of the dipole and quadrupoles in the free space. The expressions for Green's functions are provided in~\ref{section:green}.

\section{System of linear equations for calculating dipole and quadrupole moments}\label{sec:system}
Introduced in Eq.~(\ref{eq:cmm_system_matrix}) $\mathbf{Y}$ is the 24$N$ dimensional column composed of $N$ sub-vectors:
\begin{gather}
\mathbf{Y} = \left[\mathbf{Y}^1, \mathbf{Y}^2, ..., \mathbf{Y}^N \right]^T,    
\end{gather}
where $\mathbf{Y}^j$ is the sub-vector of multipole moment values corresponding to the $j$-th particle:
\begin{eqnarray}
\mathbf{Y}^j& =&\left[p_x^j, \hdots, p_z^j, m_x^j, \hdots, m_z^j, Q^j_{xx},  \hdots, Q^j_{zz},\right.\nonumber\\
&&\left. M^j_{xx}, \hdots, M^j_{zz} \right]^{T}, 
\end{eqnarray}
The vector $\mathbf{Y}_0$ is defined analogously.

The matrix $\hat{\mathbf{V}}$ is composed of blocks $\hat{V}_{jl}$ describing the interaction between multipole of the $j$-th and $l$-th particles in the array ($l \neq j$):
\begin{gather}
\hat{\mathbf{V}} = 
\begin{pmatrix}
\hat{0} & \hat{V}_{12} & \hat{V}_{13} &... &\hat{V}_{1N} \\
\hat{V}_{21} & \hat{0} & \hat{V}_{23} &... &\hat{V}_{2N} \\
\vdots  &\vdots &\vdots &\ddots &\vdots \\
\hat{V}_{N1} & \hat{V}_{N2} & \hat{V}_{N3} &... &\hat{0}
\end{pmatrix}
\end{gather}
The block $\hat{0}$ is the $24N \times 24N$ dimensional matrix of zeroes. The block $\hat{V}_{jl}$ also has a dimension of $24N \times 24N$ and consists of several sub-blocks:
\begin{gather}
\hat{V}_{jl} =
\begin{pmatrix} 
\hat{A}^{pp}_{jl} & \hat{A}^{pm}_{jl} & \hat{A}^{QQ}_{jl} & \hat{A}^{QM}_{jl} \\
\hat{B}^{pm}_{jl} & \hat{B}^{pp}_{jl} & \hat{B}^{QM}_{jl} & \hat{B}^{QQ}_{jl} \\
\hat{C}^{pp}_{jl} & \hat{C}^{pm}_{jl} & \hat{C}^{QQ}_{jl} & \hat{C}^{QM}_{jl} \\
\hat{D}^{pm}_{jl} & \hat{D}^{pp}_{jl} & \hat{D}^{QM}_{jl} & \hat{D}^{QQ}_{jl}
\end{pmatrix}.
\end{gather}
\begin{gather*}
\hat{A}^{pp}_{jl} =  \alpha_p \frac{k^2}{\varepsilon_0} \hat{G}^{pp}(\bm{r}_j, \bm{r}_l). \\  
\hat{A}^{pm}_{jl} =  \alpha_p\frac{\mathrm{i} k}{c\varepsilon_0} \hat{G}^{pm}(\bm{r}_j, \bm{r}_l). \\
\hat{A}^{QQ}_{jl} =  \alpha_p \frac{k^2}{\varepsilon_0} \hat{H}^{QQ}(\bm{r}_j, \bm{r}_l), \\
\end{gather*}
where 
\begin{gather}
\label{eq:H_QQ}
\hat{H}^{QQ}(\bm{r}_j, \bm{r}_l) = \hat{G}^{QQ}(\bm{r}_j, \bm{r}_l) \otimes \bm{n}^T_{lj}.   
\end{gather}
Here $^T$ denotes the transpose operation, $\otimes$ denotes the Kronecker product explicitly defined for (\ref{eq:H_QQ}) as:
\begin{gather*}
\begin{pmatrix}
G^{QQ}_{xx}(\bm{r}_j, \bm{r}_l) \tilde{x}_{lj}, G^{QQ}_{xx}(\bm{r}_j, \bm{r}_l) \tilde{y}_{lj},
\hdots, G^{QQ}_{xz}(\bm{r}_j, \bm{r}_l) \tilde{z}_{lj}, \\
G^{QQ}_{yx}(\bm{r}_j, \bm{r}_l) \tilde{x}_{lj}, G^{QQ}_{yx}(\bm{r}_j, \bm{r}_l) \tilde{y}_{lj},
\hdots, G^{QQ}_{yz}(\bm{r}_j, \bm{r}_l) \tilde{z}_{lj}, \\
G^{QQ}_{zx}(\bm{r}_j, \bm{r}_l) \tilde{x}_{lj}, G^{QQ}_{zx}(\bm{r}_j, \bm{r}_l) \tilde{y}_{lj},
\hdots, G^{QQ}_{zz}(\bm{r}_j, \bm{r}_l) \tilde{z}_{lj}
\end{pmatrix}
\end{gather*}
Above the components of vector $\bm{n}_{l j}$ are noted as $\bm{n}_{l j} = [\tilde{x}_{lj}, \tilde{y}_{lj}, \tilde{z}_{lj}]$, where $\tilde{x}_{lj} = (x_j - x_l)/|\bm{r}_j - \bm{r}_l|$, $\tilde{y}_{lj} = (y_j - y_l)/|\bm{r}_j - \bm{r}_l|$, $\tilde{z}_{lj} = (z_j - z_l)/|\bm{r}_j - \bm{r}_l|$.
\begin{gather*}
\hat{A}^{QM}_{jl} =  \alpha_p \frac{3 \mathrm{i} k}{c\varepsilon_0} \hat{H}^{QM}(\bm{r}_j, \bm{r}_l),
\end{gather*}
where 
\begin{gather*}
\hat{H}^{QM}(\bm{r}_j, \bm{r}_l) = \hat{G}^{QM}(\bm{r}_j, \bm{r}_l) \otimes \bm{n}^T_{lj}.   
\end{gather*}
\begin{align*}
\hat{B}^{pm}_{jl} &=  \alpha_m \frac{c k}{\mathrm{i}} \hat{G}^{pm}(\bm{r}_j, \bm{r}_l),\\
\hat{B}^{pp}_{jl} &= \alpha_m k_S^2 \hat{G}^{pp}(\bm{r}_j, \bm{r}_l).\\
\hat{B}^{QM}_{jl} &=  \alpha_m \frac{c k}{\mathrm{i}} \hat{H}^{QM}(\bm{r}_j, \bm{r}_l),\\
\hat{B}^{QQ}_{jl} &=  \alpha_m 3 k_S^2 \hat{H}^{QQ}(\bm{r}_j, \bm{r}_l).\\
\hat{C}^{pp}_{jl} &= \frac{\alpha_Q}{2} \frac{k^2}{\varepsilon_0} \hat{F}^{pp}(\bm{r}_j, \bm{r}_l),
\end{align*}
where
\begin{gather*}
\hat{F}^{pp}(\bm{r}_j, \bm{r}_l) = \hat{F}^{pp(1)}(\bm{r}_j, \bm{r}_l) + \hat{F}^{pp(2)}(\bm{r}_j, \bm{r}_l),
\end{gather*}
\begin{gather}
\label{eq:Fpp(1)_jl}
\hat{F}^{pp(1)}_{jl} = 
\begin{pmatrix}
\tfrac{\partial (G^{pp}_{jl, xx})}{\partial x} & \tfrac{\partial (G^{pp}_{jl, xy})}{\partial x} & \tfrac{\partial (G^{pp}_{jl, xz})}{\partial x} \\
\tfrac{\partial (G^{pp}_{jl, yx})}{\partial x} & \tfrac{\partial (G^{pp}_{jl, yy})}{\partial x} & \tfrac{\partial (G^{pp}_{jl, yz})}{\partial x} \\
\tfrac{\partial (G^{pp}_{jl, zx})}{\partial x} & \tfrac{\partial (G^{pp}_{jl, zy})}{\partial x} & \tfrac{\partial (G^{pp}_{jl, zz})}{\partial x} \\
\tfrac{\partial (G^{pp}_{jl, xx})}{\partial y} & \tfrac{\partial (G^{pp}_{jl, xy})}{\partial y} & \tfrac{\partial (G^{pp}_{jl, xz})}{\partial y} \\
\tfrac{\partial (G^{pp}_{jl, yx})}{\partial y} & \tfrac{\partial (G^{pp}_{jl, yy})}{\partial y} & \tfrac{\partial (G^{pp}_{jl, yz})}{\partial y} \\
\tfrac{\partial (G^{pp}_{jl, zx})}{\partial y} & \tfrac{\partial (G^{pp}_{jl, zy})}{\partial y} & \tfrac{\partial (G^{pp}_{jl, zz})}{\partial y} \\
\tfrac{\partial (G^{pp}_{jl, xx})}{\partial z} & \tfrac{\partial (G^{pp}_{jl, xy})}{\partial z} & \tfrac{\partial (G^{pp}_{jl, xz})}{\partial z} \\
\tfrac{\partial (G^{pp}_{jl, yx})}{\partial z} & \tfrac{\partial (G^{pp}_{jl, yy})}{\partial z} & \tfrac{\partial (G^{pp}_{jl, yz})}{\partial z} \\
\tfrac{\partial (G^{pp}_{jl, zx})}{\partial z} & \tfrac{\partial (G^{pp}_{jl, zy})}{\partial z} & \tfrac{\partial (G^{pp}_{jl, zz})}{\partial z} 
\end{pmatrix},
\end{gather}
\begin{gather}
\label{eq:Fpp(2)_jl}
\hat{F}^{pp(2)}_{jl} =
\begin{pmatrix}
\tfrac{\partial (G^{pp}_{jl, xx})}{\partial x} & \tfrac{\partial (G^{pp}_{jl, xy})}{\partial x} & \tfrac{\partial (G^{pp}_{jl, xz})}{\partial x} \\
\tfrac{\partial (G^{pp}_{jl, xx})}{\partial y} & \tfrac{\partial (G^{pp}_{jl, xy})}{\partial y} & \tfrac{\partial (G^{pp}_{jl, xz})}{\partial y} \\
\tfrac{\partial (G^{pp}_{jl, xx})}{\partial z} & \tfrac{\partial (G^{pp}_{jl, xy})}{\partial z} & \tfrac{\partial (G^{pp}_{jl, xz})}{\partial z} \\
\tfrac{\partial (G^{pp}_{jl, yx})}{\partial x} & \tfrac{\partial (G^{pp}_{jl, yy})}{\partial x} & \tfrac{\partial (G^{pp}_{jl, yz})}{\partial x} \\
\tfrac{\partial (G^{pp}_{jl, yx})}{\partial y} & \tfrac{\partial (G^{pp}_{jl, yy})}{\partial y} & \tfrac{\partial (G^{pp}_{jl, yz})}{\partial y} \\
\tfrac{\partial (G^{pp}_{jl, yx})}{\partial z} & \tfrac{\partial (G^{pp}_{jl, yy})}{\partial z} & \tfrac{\partial (G^{pp}_{jl, yz})}{\partial z} \\
\tfrac{\partial (G^{pp}_{jl, zx})}{\partial x} & \tfrac{\partial (G^{pp}_{jl, zy})}{\partial x} & \tfrac{\partial (G^{pp}_{jl, zz})}{\partial x} \\
\tfrac{\partial (G^{pp}_{jl, zx})}{\partial y} & \tfrac{\partial (G^{pp}_{jl, zy})}{\partial y} & \tfrac{\partial (G^{pp}_{jl, zz})}{\partial y} \\
 \tfrac{\partial (G^{pp}_{jl, zx})}{\partial z} & \tfrac{\partial (G^{pp}_{jl, zy})}{\partial z} & \tfrac{\partial (G^{pp}_{jl, zz})}{\partial z} 
\end{pmatrix},
\end{gather}
\begin{gather*}
\hat{C}^{pm}_{jl} =  \frac{\alpha_Q}{2}\frac{\mathrm{i} k}{c\varepsilon_0} \hat{F}^{pm}(\bm{r}_j, \bm{r}_l),
\end{gather*}
where 
\begin{gather*}
\hat{F}^{pm}(\bm{r}_j, \bm{r}_l) = \hat{F}^{pm(1)}(\bm{r}_j, \bm{r}_l) + \hat{F}^{pm(2)}(\bm{r}_j, \bm{r}_l).
\end{gather*}
\begin{gather*}
\hat{C}^{QQ}_{jl} =  \frac{\alpha_Q}{2} \frac{k^2}{\varepsilon_0} \hat{F}^{QQ}(\bm{r}_j, \bm{r}_l),
\end{gather*}
where
\begin{eqnarray*}
\hat{F}^{QQ}(\bm{r}_j, \bm{r}_l) =&& \left[\hat{F}^{QQ(1)}(\bm{r}_j, \bm{r}_l) + \hat{F}^{QQ(2)}(\bm{r}_j, \bm{r}_l) \right] \otimes \bm{n}^T_{lj} \\ &&+ \hat{G}^{QQ}(\bm{r}_j, \bm{r}_l) \otimes \hat{N}(\bm{r}_j, \bm{r}_l) \\ &&+ \left(\hat{G}^{QQ}(\bm{r}_j, \bm{r}_l) \otimes \hat{N}(\bm{r}_j, \bm{r}_l)\right)^T.
\end{eqnarray*}
A tensor $\hat{N}_{lj}$ is defined as following:
\begin{gather*}
\hat{N}(\bm{r}_j, \bm{r}_l) = 
\begin{pmatrix}
\dfrac{\partial \tilde{x}_{lj}}{\partial x_j}& \dfrac{\partial \tilde{x}_{lj}}{\partial y_j}& \dfrac{\partial \tilde{x}_{lj}}{\partial z_j}\\
\dfrac{\partial \tilde{y}_{lj}}{\partial x_j}& \dfrac{\partial \tilde{y}_{lj}}{\partial y_j}& \dfrac{\partial \tilde{y}_{lj}}{\partial z_j}\\
\dfrac{\partial \tilde{z}_{lj}}{\partial x_j}& \dfrac{\partial \tilde{z}_{lj}}{\partial y_j}& \dfrac{\partial \tilde{z}_{lj}}{\partial z_j}
\end{pmatrix},
\end{gather*}
where derivative of the unit vector component:
\begin{gather*}
\frac{\partial \tilde{\alpha}_{lj}}{\partial \beta_j} = \frac{\delta_{\alpha \beta} - \tilde{\alpha}_{lj}\tilde{\beta}_{lj}}{|\bm{r}_j - \bm{r}_l|},
\end{gather*}
where $\alpha = x,y,z$ and $\beta = x,y,z$.
\begin{gather*}
\hat{C}^{QM}_{jl} =  \frac{\alpha_Q}{2} \frac{3 \mathrm{i} k}{c \varepsilon_0} \hat{F}^{QM}(\bm{r}_j, \bm{r}_l),
\end{gather*}
\begin{eqnarray*}
\hat{F}^{QM}(\bm{r}_j, \bm{r}_l) &&= \\ && \left[\hat{F}^{QM(1)}(\bm{r}_j, \bm{r}_l) + \hat{F}^{QM(2)}(\bm{r}_j, \bm{r}_l) \right] \otimes \bm{n}^T_{lj} \\ &&+ \hat{G}^{QM}(\bm{r}_j, \bm{r}_l) \otimes \hat{N}(\bm{r}_j, \bm{r}_l) \\ &&+ \left(\hat{G}^{QM}(\bm{r}_j, \bm{r}_l) \otimes \hat{N}(\bm{r}_j, \bm{r}_l)\right)^T.
\end{eqnarray*}
The tensors $\hat{F}^{pm(1)}_{jl}$, $\hat{F}^{QQ(1)}_{jl}$, $\hat{F}^{QM(1)}_{jl}$ and $\hat{F}^{pm(2)}_{jl}$, $\hat{F}^{QQ(2)}_{jl}$, $\hat{F}^{QM(2)}_{jl}$ are defined as (\ref{eq:Fpp(1)_jl}) and (\ref{eq:Fpp(2)_jl}) with the corresponding Green's function, respectively.
\begin{align*}
\hat{D}^{pm}_{jl} =  \frac{\alpha_M}{2} \frac{c k}{\mathrm{i}} \hat{F}^{pm}(\bm{r}_j, \bm{r}_l),\\
\hat{D}^{pp}_{jl} =  \frac{\alpha_M}{2} k_S^2 \hat{F}^{pp}(\bm{r}_j, \bm{r}_l).\\
\hat{D}^{QM}_{jl} =  \frac{\alpha_M}{2} \frac{c k}{\mathrm{i}} \hat{F}^{QM}(\bm{r}_j, \bm{r}_l),\\
\hat{D}^{QQ}_{jl} = \frac{\alpha_M}{2} 3 k_S^2 \hat{F}^{QQ}(\bm{r}_j, \bm{r}_l).
\end{align*}

\section{Multipole dyadic Green's functions, and their derivatives}\label{section:green}
The elements of dyadic dipole and quadrupole Green’s functions for the multipole sources located in the free space~\cite{Babicheva2019}:
\begin{eqnarray}
\label{dipole_green_formula}
&&G^{pp}_{\alpha \beta}(\bm{r},\bm{r}_0) = \frac{e^{\mathrm{i} k_S l}}{4 \pi l} \left[\left(1 + \frac{\mathrm{i}}{k_S l} - \frac{1}{k_S^2 l^2}\right)\delta_{\alpha \beta}\right. \nonumber\\
&&\left. + \left(-1 - \frac{3\mathrm{i}}{k_S l} + \frac{3}{k_S^2 l^2}\right)n_{\alpha}n_{\beta}\right], \end{eqnarray}
\begin{eqnarray}
G^{pm}_{\alpha \beta}(\bm{r},\bm{r}_0) &=& -\frac{k_S e^{\mathrm{i} k_S l}}{4 \pi l} \left(\mathrm{i} - \frac{1}{k_S l}\right)\epsilon_{\alpha \beta \gamma} n_{\gamma} 
\end{eqnarray}
\begin{eqnarray}
\label{quadrupole_green_formula}
&&G^{QQ}_{\alpha \beta}(\bm{r},\bm{r}_0) = \frac{\mathrm{i} k_S e^{\mathrm{i} k l}}{24 \pi l} \times \nonumber\\ &&\left[\left(-1 - \frac{3\mathrm{i}}{k_S l} + \frac{6}{k_S^2 l^2}+\frac{6\mathrm{i}}{k_S^3 l^3}\right)\delta_{\alpha \beta} + \right. \nonumber\\ &&\left. \left(1 + \frac{6\mathrm{i}}{k_S l} - \frac{15}{k_S^2 l^2}-\frac{15\mathrm{i}}{k_S^3 l^3}\right)n_{\alpha}n_{\beta}\right], 
\end{eqnarray}
\begin{eqnarray}
\label{quadrupole_green_QM_formula}
&&G^{QM}_{\alpha \beta}(\bm{r},\bm{r}_0) = -\frac{k_S^2 e^{\mathrm{i} k_S l}}{24 \pi l} \times \nonumber\\ && \left(1 + \frac{3\mathrm{i}}{k_S l} - \frac{3}{k_S^2 l^2}\right)\epsilon_{\alpha \beta \gamma} n_{\gamma},
\end{eqnarray}
where $\bm{r} = (x,y,z)$ is the field calculation point, $\bm{r}_0=(x_0,y_0,z_0)$ is the source position, $l = |\bm{r}-\bm{r}_0|$ is the point-source distance, $\bm{n} = (\bm{r}-\bm{r}_0)/l$ is the unit vector, $\delta_{\alpha \beta}$ is the Kronecker symbol. Greek letters $\alpha, \beta, \gamma$ denote the Cartesian coordinates $x,y,z$. 

Elements of the Cartesian derivatives of dyadic Green's functions (\ref{dipole_green_formula})-(\ref{quadrupole_green_QM_formula}):
\begin{eqnarray}
\label{eq:d_Gpp}
&&\frac{\partial}{\partial \gamma} G^{pp}_{\alpha \beta}(\bm{r},\bm{r}_0) = \frac{k_S^2 e^{\mathrm{i} k_S l}}{4 \pi} \times \nonumber\\ &&\left\{ \left(\frac{\mathrm{i}}{k_S l} - \frac{2}{k_S^2 l^2}-\frac{3\mathrm{i}}{k_S^3 l^3} + \frac{3}{k_S^4 l^4}\right)\delta_{\alpha \beta} n_{\gamma} \right. \nonumber\\ &&\left. +\left(-\frac{1}{k_S^2 l^2} - \frac{3\mathrm{i}}{k_S^3 l^3} + \frac{3}{k_S^4 l^4}\right)\left(\delta_{\alpha \gamma} n_{\beta}+ \delta_{\beta \gamma} n_{\alpha} \right)
\right. \nonumber\\ &&\left. + \left(-\frac{\mathrm{i}}{k_S l} + \frac{6}{k_S^2 l^2}+\frac{15\mathrm{i}}{k_S^3 l^3} - \frac{15}{k_S^4 l^4}\right)n_{\alpha}n_{\beta}n_{\gamma}\right\},
\end{eqnarray}
\begin{eqnarray}
&&\frac{\partial }{\partial \gamma} G^{pm}_{\alpha \beta}(\bm{r},\bm{r}_0) = -\frac{k_S^3 e^{\mathrm{i} k_S l}}{4 \pi} \epsilon_{\alpha \beta \tau} \left\{\left(\dfrac{\mathrm{i}}{k_S^2 l^2} - \dfrac{1}{k_S^3 l^3}\right) \delta_{\tau \gamma} \right. \nonumber\\ &&\left. + \left(- \dfrac{1}{k_S l}-\dfrac{3\mathrm{i}}{k_S^2 l^2} + \dfrac{3}{k_S^3 l^3}\right) n_{\tau}n_{\gamma}\right\},
\end{eqnarray}
\begin{eqnarray}
\label{eq:d_GQQ}
&&\frac{\partial }{\partial \gamma} G^{QQ}_{\alpha \beta}(\bm{r},\bm{r}_0) = \frac{\mathrm{i} k_S^2 e^{\mathrm{i} k_S l}}{24 \pi l} \times \nonumber\\ &&\left\{   \left(-\mathrm{i} + \dfrac{4}{k_S l}+\dfrac{12\mathrm{i}}{k_S^2 l^2} - \dfrac{24}{k_S^3 l^3}-\dfrac{24\mathrm{i}}{k_S^4 l^4}\right) \delta_{\alpha \beta} n_{\gamma} \right. \nonumber\\ &&+ \left(\dfrac{1}{k_S l} + \dfrac{6\mathrm{i}}{k_S^2 l^2} - \dfrac{15}{k_S^3 l^3}-\dfrac{15\mathrm{i}}{k_S^4 l^4}\right)\left(\delta_{\alpha \gamma} n_{\beta}+\delta_{\beta \gamma} n_{\alpha}\right)
\nonumber\\ &&\left. + \left(\mathrm{i} - \dfrac{9}{k_S l}-\dfrac{39\mathrm{i}}{k_S^2 l^2} + \dfrac{90}{k_S^3 l^3}+\dfrac{90\mathrm{i}}{k_S^4 l^4}\right)n_{\alpha}n_{\beta}n_{\gamma}\right\},
\end{eqnarray}
\begin{eqnarray}
\label{eq:d_GQM}
&&\frac{\partial }{\partial \gamma}G^{QM}_{\alpha \beta}(\bm{r},\bm{r}_0) = -\frac{k_S^3 e^{\mathrm{i} k_S l}}{24 \pi l} \epsilon_{\alpha \beta \tau}\times \nonumber\\ && \left\{\left(\frac{1}{k_S l} + \dfrac{3 \mathrm{i}}{k_S^2 l^2} - \dfrac{3}{k_S^3 l^3}\right) \delta_{\tau \gamma} \right. \nonumber \\ &&\left. + \left(\mathrm{i} - \dfrac{5}{k_S l}-\dfrac{12\mathrm{i}}{k_S^2 l^2} + \dfrac{12}{k_S^3 l^3}\right) n_{\tau}n_{\gamma}\right\}.
\end{eqnarray}

\section{Solution of CMM equation for nanoparticle dimer at the wavelengths of isolated MD, EQ and MQ resonances}\label{sec:solutions}
The solution of Eq. (\ref{eq:cmm_system_matrix}) in the framework of specific resonant MD response for the dimer placed in free space ($\varepsilon_S = 1$, $k = k_S$):
\begin{gather*}
    m^j_{\beta} = \frac{\alpha_m H_0}{1 - \alpha_m k^2 G^{pp}_{12, \beta\beta}}, \quad j=1,2.
\end{gather*}
where $\beta = y$ for $\bm{E}_0 \parallel x$, and $\beta = x$ for $\bm{E}_0 \parallel y$. 

The solution for isolated EQ resonance:
\begin{gather*}
    Q^j_{\beta z} = \frac{\frac{\alpha_Q}{2}\mathrm{i}k E_0}{1 - \frac{\alpha_Q}{2} k^2 \varepsilon_0^{-1} B^{QQ}_{12, \beta z}}, \quad j=1,2,
\end{gather*}
where $\beta = x$ for $\bm{E}_0 \parallel x$, and $\beta = y$ for $\bm{E}_0 \parallel y$. The expression for $B^{QQ}_{12, \beta z}$ are followed from the quadrupole Green's tensors (\ref{quadrupole_green_formula}), and its derivative (\ref{eq:d_GQQ}):
\begin{eqnarray}
\label{eq:definition_BQQ}
&B^{QQ}_{12, \beta z} = \\
&\begin{cases}
\frac{\mathrm{i} k^3}{24 \pi} \frac{e^{\mathrm{i} k D}}{k D} \left[-\mathrm{i} + \frac{5}{k D} + \frac{21 \mathrm{i}}{k^2 D^2} - \frac{48}{k^3 D^3} - \frac{48}{k^4 D^4} \right],& \beta = y \nonumber\\
\frac{\mathrm{i} k^3}{12 \pi} \frac{e^{\mathrm{i} k D}}{k^2 D^2} \left[-1 - \frac{3\mathrm{i}}{k D} + \frac{6}{k^2 D^2}+\frac{6\mathrm{i}}{k^3 D^3}\right],& \beta = x
\end{cases}.
\end{eqnarray}

The solution for isolated MQ resonance:
\begin{gather*}
    M^j_{\beta z} = \frac{\frac{\alpha_M}{2}\mathrm{i}k H_0}{1 - \frac{\alpha_M}{2} 3 k^2 B^{QQ}_{12, \beta z}}, \quad j=1,2,
\end{gather*}
where $\beta = y$ for $\bm{E}_0 \parallel x$, and $\beta = x$ for $\bm{E}_0 \parallel y$.

\section{Ring dipole sum}\label{appendix:ring_dipole_sum}
In order to find the fields generated at the position of the $j$-th dipole by all other radially oriented dipole scatterers formed a ring of radius $R$, one need to deal with the sum:
\begin{gather}
\label{eq:G0}
    \tilde{G}^{pp}_{\rho \rho} = \sum\limits_{l=1, l \ne j}^{l = N} G^{pp}_{\rho \rho}(\bm{r}_j, \bm{r}_l),
\end{gather}
where $\bm{r}_l = R[\cos(\varphi_l), \sin(\varphi_l), 0]^T$, and $G^{pp}_{\rho \rho}$ is the component of the vacuum dipole dyadic Green's function in cylindrical coordinates given by the following expression:
\begin{multline}
    G^{pp}_{\rho \rho}(\bm{r}_j, \bm{r}_l) = \dfrac{k e^{\mathrm{i} k D_{jl}}}{4 \pi} \left\{\left( \frac{1}{k D_{jl}} + \frac{\mathrm{i}}{k^2 D_{jl}^2} - \frac{1}{k^3 D_{jl}^3} \right) \right. \\ \left. +\sin^2 \left( \frac{\varphi_j - \varphi_l}{2} \right) \left( - \frac{1}{k D_{jl}} + \frac{\mathrm{i}}{k^2 D_{jl}^2} - \frac{1}{k^3 D^3_{jl}} \right)\right\},
\label{GFCYL2}
\end{multline}
where $D_{jl} = 2 R \sin\left( \frac{\left|\varphi_j - \varphi_l\right|}{2} \right)$ is the distance between the dipoles $j$, and $l$. 

Let us consider the limit of an infinitely large ring so that $N, R \to \infty$, but simultaneously, the distance between the neighboring dipoles $D = 2 R \sin \left( \frac{\pi}{N} \right)$ is kept fixed. First of all, notice that $\lim\limits_{N \to \infty} D_{j l} = \lim\limits_{N \to \infty} 2 R \sin\left( \frac{\left| \varphi_j - \varphi_l \right|}{2} \right) = \lim\limits_{N \to \infty} 2 R \sin \left( \frac{\pi}{N} \right) \frac{ \sin\left( \frac{\pi \left| j - l \right|}{N} \right)}{\sin \left( \frac{\pi}{N} \right)} = D \left| j - l \right|$. Secondly, in Eq.~(\ref{GFCYL2}), the term $\sin^2 \left( \frac{\varphi_j - \varphi_l}{2} \right)$ vanishes as $N \to \infty$, and we are left with:
\begin{eqnarray}
    &&\lim\limits_{N,R \to \infty} \tilde{G}^{pp}_{\rho \rho} = \sum\limits_{{}_{\: \: l \ne j}^{l = -\infty}}^{\infty} \lim\limits_{N,R \to \infty} G^{pp}_{\rho \rho}(\bm{r}_j, \bm{r}_l) = \\ &&\sum\limits_{{}_{\: \: l \ne j}^{l = -\infty}}^{\infty} \tfrac{k e^{\mathrm{i} k D |j-l|}}{4 \pi} \left( \tfrac{1}{k D |j-l|} + \tfrac{\mathrm{i}}{\left( k D |j-l| \right)^2} - \tfrac{1}{ \left( k D |j-l|\right)^3} \right), \nonumber
\end{eqnarray}
which is exactly the lattice sum for an infinite periodic chain of transversally oriented dipole moments,  which is known analytically~
\cite{Citrin2006}. Finally, this allows us to write the following:
\begin{gather}
    \dfrac{\mathrm{i} 6 \pi}{k} \tilde{G}^{pp}_{\rho \rho} \to 3 \left( \dfrac{\mathrm{i} \text{Li}_1(e^{\mathrm{i} k D})}{k D} - \dfrac{\text{Li}_2(e^{\mathrm{i} k D})}{\left(k D\right)^2} - \dfrac{\mathrm{i} \text{Li}_3(e^{\mathrm{i} k D})}{\left(k D\right)^3} \right),
\end{gather}
where $\text{Li}_s(e^{\mathrm{i} k D})$ is the polylogarithm function of order $s$ ($s = 1,2,3$) and argument $e^{\mathrm{i} k D}$. 

\section*{References}
\bibliographystyle{iopart-num}
\bibliography{draft_iop}
\end{document}